\documentclass{amsart}
\usepackage{graphicx,amssymb}

\theoremstyle{definition}

\theoremstyle{remark}

\def\beq{\begin{eqnarray}}
\def\eeq{\end{eqnarray}}

\def\Ri2{R_{\mu\nu}R^{\mu\nu}}

\begin{document}
\title{On the evolution of universes in quadratic theories of gravity}
\author[J.D. Barrow and S. Hervik]{John D. Barrow$^{1}$ and Sigbj{\o }rn
Hervik$^{2}$}
\address{$^{1}$DAMTP, Centre for Mathematical Sciences, Cambridge
University, Wilberforce Rd., Cambridge CB3 0WA, UK\\
$^{2}$Department of Mathematics \& Statistics, Dalhousie University,
Halifax, Nova Scotia, Canada B3H 3J5 }
\email{herviks@mathstat.dal.ca\\
J.D.Barrow@damtp.cam.ac.uk}
\date{\today }

\begin{abstract}
We use a dynamical systems approach to investigate Bianchi type I and II
universes in quadratic theories of gravity. Due to the complicated nature of
the equations of motion we focus on the stability of exact solutions and
find that there exists an isotropic FRW universe acting as a past attractor.
This may indicate that there is an isotropisation mechanism at early times
for these kind of theories. We also discuss the Kasner universes, elucidate
the associated centre manifold structure, and show that there exists a set
of non-zero measure which has the Kasner solutions as a past attractor.
Regarding the late-time behaviour, the stability shows a dependence of the
parameters of the theory. We give the conditions under which the de Sitter
solution is stable and also show that for certain values of the parameters
there is a possible late-time behaviour with phantom-like behaviour. New
types of anisotropic inflationary behaviour are found which do not have
counterparts in general relativity.
\end{abstract}

\maketitle

\section{Introduction}

In this paper we are going to study the dynamical evolution of a class of
anisotropic universes in quadratic theories of gravity. These extensions of
general relativity (GR) provide guidance as to the possible effects of
quantum corrections to the Einstein equations. They permit us to investigate
the possible effects on singularity formation, inflation, and the expansion
dynamics of the early universe. Past studies of these extensions have
focused on the isotropic Friedmann metrics, where it is sufficient to
consider only the effects of an $R^{2}$ term in the gravitational lagrangian
to the field equations \cite{R2}. However, the $R^{2}$ contribution has
fairly predictable cosmological consequences because the resulting quadratic
vacuum theory is conformally equivalent to GR with a scalar field moving in
a potential with a single asymmetric minimum \cite{barrcot, barr}. This type
of solution has been well studied in connection with inflation \cite{Inf}
and in the situation of pure power-law lagrangians of the form $R^{n}$, \cite%
{Schmidt, BC, BC1, BC2}. The addition of the $\Ri2$ Ricci term to the
lagrangian in the case of an anisotropic universe creates a much richer
diversity of cosmological behaviours that are harder to summarise in terms
of simple modifications of the general-relativistic situation. The effective
stresses that are contributed to the field equations by the quadratic Ricci
terms in the lagrangian can mimic a wide range of fluids which violate the
strong and weak energy conditions. This allows completely different
behaviour to occur than is found in GR or its quadratic extension with pure $%
R^{2}$ contributions. In particular, we have shown elsewhere that the cosmic
no-hair theorems no longer hold: vacuum universes with positive cosmological
constant do not necessarily approach de Sitter, but can inflate
anisotropically \cite{BHII}. Moreover, the addition of quadratic Ricci terms
to the lagrangian can create cosmological models which have no counterpart
in the GR limit of the theory.

In what follows, we are going to widen this investigation by considering the
global dynamics of the Bianchi type I and II universes in the presence of
quadratic Ricci and scalar curvature contributions to the lagrangian. \
Whilst these anisotropic universes contain some mathematical
simplifications, they include spatially homogeneous expanding universes with
isotropic (type I) and anisotropic (type II) spatial 3-curvatures as well as
expansion shear anisotropy. To this end, we consider the gravitational
action 
\begin{equation}
S_{G}=\frac{1}{2\kappa }\int_{M}\mathrm{d}^{4}x\sqrt{|g|}\left( R+\alpha
R^{2}+\beta \Ri2-2\Lambda \right) .  \label{a}
\end{equation}%
Variation of this action leads to the following generalised Einstein
equations (see, e.g., \cite{DT}): 
\begin{equation}
G_{\mu \nu }+\Phi _{\mu \nu }+\Lambda g_{\mu \nu }=\kappa T_{\mu \nu },
\label{field}
\end{equation}%
where $T_{\mu \nu }$ is the energy-momentum tensor of the ordinary matter
sources, which in this paper we will assume to be zero, for simplicity, 
\begin{eqnarray}  \label{c}
G_{\mu \nu } &\equiv &R_{\mu \nu }-\frac{1}{2}Rg_{\mu \nu },  \label{b} \\
\Phi _{\mu \nu } &\equiv &2\alpha R\left( R_{\mu \nu }-\frac{1}{4}Rg_{\mu
\nu }\right) +(2\alpha +\beta )\left( g_{\mu \nu }\Box -\nabla _{\mu }\nabla
_{\nu }\right) R  \notag \\
&&+\beta \Box \left( R_{\mu \nu }-\frac{1}{2}Rg_{\mu \nu }\right) +2\beta
\left( R_{\mu \sigma \nu \rho }-\frac{1}{4}g_{\mu \nu }R_{\sigma \rho
}\right) R^{\sigma \rho },
\end{eqnarray}%
with $\Box \equiv \nabla ^{\mu }\nabla _{\mu }$ and $g_{\mu \nu }$ is the
metric tensor and $\Lambda $ \ the cosmological constant. The effective
stress tensor $\Phi _{\mu \nu }$ incorporates the deviation from regular
Einstein gravity introduced by the quadratic terms in the action, and we see
that $\alpha =\beta =0$ implies $\Phi _{\mu \nu }=0,$although the converse
need not be true.

Similarly, by rescaling $(\alpha ,\beta )\mapsto \kappa (\alpha ,\beta )$ we
can consider the limit $\kappa \rightarrow \infty $. This limit, whenever it
exists, corresponds to the case where the action only contains quadratic
terms, hence eq. (\ref{field}) reduces to (for vacuum, $T_{\mu \nu }=0$) 
\begin{equation}
\Phi _{\mu \nu }=0.  \label{Phi=0}
\end{equation}%
It is important to consider the vacuum solutions to the pure quadratic
theories; that is, solutions to eq.(\ref{Phi=0}). First, we note that any
Einstein metric with $R_{\mu \nu }=\lambda g_{\mu \nu }$, necessarily obeys $%
\Phi _{\mu \nu }=0$. In particular, this implies that \emph{any GR vacuum
solution ($G_{\mu \nu }=0$) will also be a vacuum solution to the quadratic
theory ($G_{\mu \nu }+\Phi _{\mu \nu }=0$).} This means that there will be
chaotic dynamics in the quadratic theory because it is present in a number
of general-relativistic vacuum Bianchi models (types VI$_{-1/9}^{\ast }$,
VIII, and IX) near an initial curvature singularity; however it might not be
stable in solutions to the quadratic theories in the way that it appears to
be for these solutions in GR. Second, there are some specific solutions to $%
\Phi _{\mu \nu }=0$, which are not solutions to $G_{\mu \nu }=0,$ and so
have no counterparts in the $(\alpha ,\beta )\rightarrow (0,0)$
general-relativity limit of the quadratic theory. This feature plays an
important role in the evolution of universes that arise in these theories.
For example, one such intrinsically quadratic solution is a
Friedmann-Robertson-Walker (FRW) universe which has expansion dynamics
similar to those of a radiation-dominated universe in GR. This means that if
we take a FRW universe with spatial curvature parameter $k$, there is a
simple solution to $\Phi _{\mu \nu }=0$ given by the metric that solves the
field equations of GR $(G_{\mu \nu }=\kappa T_{\mu \nu })$ in the presence
of black-body radiation: 
\begin{equation}
\mathrm{d}s^{2}=-\mathrm{d}t^{2}+(t-kt^{2})\left[ \frac{\mathrm{d}r^{2}}{%
1-kr^{2}}+r^{2}(\mathrm{d}\phi ^{2}+\sin ^{2}\phi \mathrm{d}\theta ^{2})%
\right] .  \label{quadrad}
\end{equation}%
Note that despite the fact that $T_{\mu \nu }=0$ this quadratic universe
behaves as if it is radiation-dominated. This radiation-like universe seems
to play a special role in the theory as it will be found to act as a past
attractor for the anisotropic universes we study. Other interesting
conclusions can also be deduced from this simple solution. We see, for
example, that unlike for the GR case, in the quadratic theory there is a
closed ($k=1$) FRW \emph{vacuum} solution which recollapses after a finite
time $(t=1)$ and a flat $(k=0)$ vacuum solution.

There are some further geometrical observations regarding solutions to these
quadratic theories that will be useful. In 4D spacetimes we can use the Weyl
curvature invariant and the Euler density, $E$, defined by, 
\begin{eqnarray}
C_{\mu \nu \rho \sigma }C^{\mu \nu \rho \sigma } &=&R_{\mu \nu \rho \sigma
}R^{\mu \nu \rho \sigma }-2R_{\mu \nu }R^{\mu \nu }+\frac{1}{3}R^{2},  \notag
\\
E &=&R_{\mu \nu \rho \sigma }R^{\mu \nu \rho \sigma }-4R_{\mu \nu }R^{\mu
\nu }+R^{2},
\end{eqnarray}%
to replace the quadratic Ricci invariant in the action, since 
\begin{equation*}
\alpha R^{2}+\beta R_{\mu \nu }R^{\mu \nu }=\tfrac{1}{3}(3\alpha +\beta
)R^{2}+\tfrac{\beta }{2}\left( C_{\mu \nu \rho \sigma }C^{\mu \nu \rho
\sigma }-E\right) .
\end{equation*}%
Since integration over the Euler density is a topological invariant, the
variation of $E$ will not contribute to the equations of motion. From this
we see that there are two special cases

\begin{itemize}
\item[a.] $(3\alpha +\beta )=0$ : the conformally invariant Bach-Weyl
gravity theory \cite{bach}.

\item[b.] $\beta =0$ : A special case of $f(R)$ gravity.
\end{itemize}

Since the equations of motion change their structure in these two special
cases, a separate analysis is required for each. So, in what follows we will
assume that 
\begin{equation*}
3\alpha +\beta \neq 0,\quad \beta \neq 0.
\end{equation*}%
On the other hand, our formulation of the equations of motion is well
defined in the $\kappa \rightarrow \infty $ limit (as explained above).

In our analysis, we will adopt the dynamical systems approach by introducing
expansion-normalised variables. This approach has proven to be extremely
successful in the analysis of spatially homogeneous Bianchi universes in GR 
\cite{DS1,tilt}. However, by considering quadratic theories of gravity there
are some extra complications added to the formalism, as the universe can
bounce or recollapse at expansion minima and maxima, which leads to
infinities in the expansion-normalised variables. Here, we will discuss
these infinities and explain how a sequence of expanding and contracting
phases can be considered. On the other hand, initial singularities and
future asymptotes are all finite in these variables, and the dynamical
systems approach is ideal for studying such regimes.

In a recent paper by Leach \textit{et al} \cite{LCD}, a dynamical systems
approach to the local rotational symmetric Bianchi type I universes was
considered within a class of $f(R)$ theories. In particular, they were
interested in the shear dynamics of such universes, and, interestingly, they
found that there is a possibility for an isotropic singularity. The
introduction of a $\Ri2$ term introduces extra shear degrees of freedom and
their higher time derivatives, which considerately complicate the equations
of motion. However, in spite of the fact that the case we study is is a
quite separate, and we consider more general Bianchi models, we also find
the possibility for an isotropic singularity, as did Cotsakis et al \cite%
{cotsakis} in their study of Bianchi type IX universes in quadratic
theories. Indeed, we will argue that this isotropic singularity is past
stable for \emph{all Bianchi models}.

\section{Equations of motion}

Our starting point is the generalised vacuum field equations: 
\begin{equation*}
G_{\mu \nu }+\Phi _{\mu \nu }+\Lambda g_{\mu \nu }=0.
\end{equation*}%
We note that the GR-limit can be obtained by letting $(\alpha ,\beta
)\rightarrow (0,0)$. Now consider the spatially homogeneous Bianchi metrics.
We can always write their metric line-elements as 
\begin{equation*}
\mathrm{d}s^{2}=-\mathrm{d}t^{2}+\delta _{ab}{\mbox{\boldmath${\omega}$}}^{a}%
{\mbox{\boldmath${\omega}$}}^{b},
\end{equation*}%
where ${\mbox{\boldmath${\omega}$}}^{a}$ is a triad of one-forms obeying 
\begin{equation*}
\left( \mathrm{d}{\mbox{\boldmath${\omega}$}}^{a}\right) _{\perp }=-\frac{1}{%
2}C_{~bc}^{a}{\mbox{\boldmath${\omega}$}}^{b}\wedge {\mbox{\boldmath${%
\omega}$}}^{c},
\end{equation*}%
the $C_{~bc}^{a}$ depend only on time and at any moment of constant time are
the structure constants of the Bianchi group-type under consideration, and $%
(-)_{\perp }$ means projection onto the spatially-homogeneous hypersurfaces.
The structure constants can be split into a vector part $a_{b}$ and a
symmetric tensor $n^{ab}$ in the standard way by 
\begin{equation*}
C_{~bc}^{a}=\varepsilon _{bcd}n^{da}-\delta _{~b}^{a}a_{c}+\delta
_{~c}^{a}a_{b}.
\end{equation*}%
Using the Jacobi identity, $a_{b}$ is in the kernel of $n^{ab}$ 
\begin{equation*}
n^{ab}a_{b}=0.
\end{equation*}%
The different structures of $n^{ab}$ and $a_{b}$ define the various Bianchi
types (see, e.g., \cite{DS1}).

Defining the time-like hypersurface-orthogonal vector $\mathbf{u}=\partial
/\partial t,$ we can define the Hubble scalar, $H$, and the shear, $\sigma
_{ab}$, as follows: 
\begin{equation*}
H\equiv \frac{1}{3}u_{;\mu }^{\mu },\quad \sigma _{ab}=u_{(a;b)}-H\delta
_{ab}.
\end{equation*}%
We will also restrict attention to cosmological models where the shear is
diagonal, so we can write 
\begin{equation*}
\sigma _{ab}=\text{diag}(-2\sigma _{+},\sigma _{+}+\sqrt{3}\sigma
_{-},\sigma _{+}-\sqrt{3}\sigma _{-}).
\end{equation*}

We are interested in the Bianchi type I and type II models for which $a_b=0$
and we can write 
\begin{eqnarray}
n_{ab}=0 \quad \text{(type I),} \qquad n_{ab}=\mathrm{diag}(n_{11},0,0)
\quad \text{(type II)}.  \notag
\end{eqnarray}

We define the dimensionless expansion-normalised variables by scaling out
appropriate powers of $H$%
\begin{eqnarray}
&&B=\frac{1}{(3\alpha +\beta )H^{2}},\quad \chi =\frac{\beta }{3\alpha
+\beta },  \notag \\
&&Q=\frac{\dot{H}}{H^{2}},\quad Q_{2}=\frac{\ddot{H}}{H^{3}},\quad \Omega
_{\Lambda }=\frac{\Lambda }{3H^{2}},\quad N=\frac{n_{11}}{\sqrt{3}H},  \notag
\\
&&\Sigma _{\pm }=\frac{\sigma _{\pm }}{H},\quad \Sigma _{\pm 1}=\frac{\dot{%
\sigma}_{\pm }}{H^{2}},\quad \Sigma _{\pm 2}=\frac{\ddot{\sigma}_{\pm }}{%
H^{3}}.
\end{eqnarray}%
Note the presence of time derivatives of the variables $Q_{2}$ and $\Sigma
_{\pm 2};$ this reflects the $4^{th}$-order time derivatives in the field
equations of the quadratic theory\footnote{%
The variable $Q_{2}$ can be related to the statefinders $q$ and $j$ (see
e.g. \cite{Dab}), by $Q_{2}=j+3q+2$.}; $\chi $ is a constant. We also
introduce the dynamical time variable $\tau $ by 
\begin{equation*}
\frac{\mathrm{d}\tau }{\mathrm{d}t}=H,
\end{equation*}%
and we assume that the cosmological constant is positive: $\Omega _{\Lambda
}>0$.

The equations of motion and the constraint ('Friedmann-like') equation are: 
\begin{eqnarray}
B^{\prime } &=&-2QB, \\
\Omega _{\Lambda }^{\prime } &=&-2Q\Omega _{\Lambda }, \\
N^{\prime } &=&-(Q+1+4\Sigma _{+})N, \\
Q^{\prime } &=&-2Q^{2}+Q_{2}, \\
Q_{2}^{\prime } &=&-3(Q+2)Q_{2}-\frac{9}{2}(Q+2)Q-\frac{3}{4}B\left(
1+\Sigma ^{2}-\Omega _{\Lambda }+\frac{2}{3}Q-\frac{1}{3}N^{2}\right)  \notag
\\
&&-\frac{3}{2}(1+2\chi )\Sigma ^{4}-\frac{1}{4}(8+\chi )\Sigma
_{1}^{2}-(4-\chi )(\Sigma \cdot \Sigma _{1})  \notag \\
&&-\frac{1}{4}(4-\chi )(3\Sigma ^{2}+2\Sigma \cdot \Sigma _{2}+2Q\Sigma
^{2})-(1+2Q)N^{2}  \notag \\
&&+N^{2}\left[ \frac{1}{2}(1+8\chi )N^{2}+5(13+3\chi )\Sigma
_{+}^{2}+8(2\Sigma _{+}-\Sigma _{+1})+(1-\chi )\Sigma _{-}^{2}\right] , \\
\Sigma _{\pm }^{\prime } &=&-Q\Sigma _{\pm }+\Sigma _{\pm 1}, \\
\Sigma _{\pm 1}^{\prime } &=&-2Q\Sigma _{\pm 1}+\Sigma _{\pm 2}, \\
\Sigma _{+2}^{\prime } &=&-3(Q+2)\Sigma _{+2}+\frac{\Sigma _{+1}}{\chi }%
\left[ B-(11\chi -8)+4Q(1-\chi )+4\Sigma ^{2}(1+2\chi )\right]  \notag \\
&&+\frac{\Sigma _{+}}{\chi }\left[ 3B+(4-\chi )(6+Q_{2}+7Q)+4(1+2\chi
)(3\Sigma ^{2}+2\Sigma \cdot \Sigma _{1})\right]  \notag \\
&&-\frac{4}{\chi }N^{2}\left[ B+8+4Q-4(1+8\chi )N^{2}\right]  \notag \\
&&-\frac{4}{\chi }N^{2}\left[ (1+15\chi )(\Sigma _{+}+\Sigma _{+1}-4\Sigma
_{+}^{2})+4(1-\chi )\Sigma _{-}^{2}\right]  \label{eqSigma+2} \\
\Sigma _{-2}^{\prime } &=&-3(Q+2)\Sigma _{-2}+\frac{\Sigma _{-1}}{\chi }%
\left[ B-(11\chi -8)+4Q(1-\chi )+4\Sigma ^{2}(1+2\chi )\right]  \notag \\
&&+\frac{\Sigma _{-}}{\chi }\left[ 3B+(4-\chi )(6+Q_{2}+7Q)+4(1+2\chi
)(3\Sigma ^{2}+2\Sigma \cdot \Sigma _{1})\right]  \notag \\
&&-\frac{4(1-\chi )}{\chi }N^{2}(\Sigma _{-}+\Sigma _{-1}-8\Sigma _{-}\Sigma
_{+}).  \label{eqSigma-2}
\end{eqnarray}%
These equations are subject to the constraint: 
\begin{eqnarray}
0 &=&B(1-\Omega _{\Lambda }-\Sigma ^{2}-N^{2})+12Q-2Q^{2}+4Q_{2}-(4-\chi
)(3+2Q)\Sigma ^{2}  \notag \\
&&-6(1+2\chi )\Sigma ^{4}-\chi (\Sigma _{1}^{2}-2\Sigma \cdot \Sigma
_{2})+4(2+\chi )(\Sigma \cdot \Sigma _{1})  \notag \\
&&+4N^{2}\left[ \frac{1}{2}(1+8\chi )N^{2}+1+(1+15\chi )\Sigma
_{+}^{2}+8\Sigma _{+}+(1-\chi )\Sigma _{-}^{2}\right] ,
\end{eqnarray}%
where, we have introduced the short-hand notation $\Sigma _{n}\equiv (\Sigma
_{+n},\Sigma _{-n})$ and $(\Sigma _{n}\cdot \Sigma _{m})\equiv \Sigma
_{+n}\Sigma _{+m}+\Sigma _{-n}\Sigma _{-m}$.

There are two points to note. First, the parameter $Q$ is related to the
usual deceleration parameter $q$ via 
\begin{equation*}
q=-(1+Q).
\end{equation*}%
Second, the variable $B$ measures how greatly the quadratic part of the
lagrangian dominates over the general-relativistic Einstein-Hilbert term $%
R-2\Lambda $. In particular, the larger the value of $B,$ the "closer" we
are to GR. The $B=0$ case corresponds to a purely quadratic lagrangian
theory whose equations of motion reduce to $\Phi _{\mu \nu }=0$.

The equations of motion define a dynamical flow in a large phase volume, and
the behaviour can be analysed qualitatively by standard techniques from the
theory of ordinary differential equations. Of particular interest are the
exact solutions which define the critical points of the system and their
stability at small and large times.

\section{Solutions and their stability}

\subsection{The de Sitter solution: $\mathrm{dS}$}

The de Sitter solution is characterised by the critical points where 
\begin{equation*}
Q=Q_{2}=\Sigma _{\pm }=\Sigma _{\pm 1}=\Sigma _{\pm 2}=N=0,\quad \Omega
_{\Lambda }=1,\quad B\neq 0.
\end{equation*}%
Its stability is determined by the eigenvalues:\footnote{%
The stability of de Sitter for these models was also studied independently
by A. Toporensky and collaborators \cite{topMG11}.} 
\begin{equation*}
0,~-1,~-\frac{3}{2}\left( 1\pm \sqrt{1-\frac{2B}{9}}\right) ,~-3[\times 2],~-%
\frac{3}{2}\left( 1\pm \sqrt{1+\frac{4[B+2(4-\chi )]}{9\chi }}\right)
[\times 2]
\end{equation*}%
The zero eigenvalue corresponds to the variable $B$ and appears because
these solutions are a one-parameter family. The last three eigenvalues arise
from the shear equations and come therefore in pairs. We also see that the
limit $\chi \rightarrow 0$ is ill-defined for these eigenvalues and these
arise therefore only for theories with a non-zero $\Ri2$-term\footnote{%
As can be seen from the equations of motion, eqs. (\ref{eqSigma+2}) and (\ref%
{eqSigma-2}) are ill-defined in the $\chi \rightarrow 0$ limit. The case $%
\chi =0$ induces a relation between $\Sigma _{1}$ and the other variables
which makes it possible to eliminate $\Sigma _{2}$ and $\Sigma _{1}$ from
the equations of motion. Therefore the case $\chi =0$ has 4 fewer variables,
which correspond to the last two pairs of eigenvalues given above.}.

We note that 
\begin{equation*}
B>0\Rightarrow (3\alpha +\beta )>0,\quad \frac{B+2(4-\chi )}{\chi }%
<0\Rightarrow \frac{1+2\Lambda (4\alpha +\beta )}{\beta }<0.
\end{equation*}%
In particular, this means that whenever $\alpha =0$, $\beta >0$ the de
Sitter solution has two unstable modes. These modes stem from the shear
equations and are therefore not present for the FRW models studied earlier.
On the other hand, if $\beta <0$ and $(3\alpha +\beta )>0$ then the de
Sitter solution is stable.

A related set of equilibrium points is 
\begin{equation*}
Q=Q_{2}=\Sigma _{\pm }=\Sigma _{\pm 1}=\Sigma _{\pm 2}=N=B=0,\quad \Omega
_{\Lambda }=\mathrm{constant}\neq 0.
\end{equation*}%
The eigenvalues for this line bifurcation is the same as for the de Sitter
solution restricted to $B=0$. Consequently, there exist two zero eigenvalues
for this set of solutions. However, it can be shown that these solutions are
always unstable.

\subsection{The Kasner circle: $\mathcal{K}$}

The Kasner circle of equilibrium points is given by 
\begin{eqnarray}
&&Q=-3,\quad (\Sigma _{+},\Sigma _{-})=(\cos \phi ,\sin \phi ),\quad (\Sigma
_{+1},\Sigma _{-1})=-3(\cos \phi ,\sin \phi ),  \notag \\
&&(\Sigma _{+2},\Sigma _{-2})=18(\cos \phi ,\sin \phi ),\quad \Omega
_{\Lambda }=B=N=0.
\end{eqnarray}%
The exact Kasner solutions are completed by integrating the $B$ equation so
that $B=B_{0}e^{6\tau }$; however, here we will mostly be interested in the
limit $\tau \rightarrow -\infty $ for which case the equilibrium points
suffice.

Stability is determined by the eigenvalues: 
\begin{equation*}
0[\times 2],\quad 2(1-2\cos \phi ),\quad 6[\times 7].
\end{equation*}%
We note that there are two zero eigenvalues. One of the zero eigenvalues
just corresponds to the fact that this is a line of non-isolated equilibria.
The second eigenvalue corresponds to a one-dimensional centre manifold and
in order to determine the stability of the Kasner circle an analysis of this
centre manifold is needed. Note also that the eigenvalue $2(1-2\cos \phi )$
arises from the curvature variable $N$. This is the same mode as in the GR
case, where it causes vacuum type II transitions between points on the
Kasner circle (see e.g. \cite{DS1}).

\paragraph{\textbf{The centre manifold}}

Consider the invariant subspace given by $(Q,\Sigma,F,G)$ defined as
follows: 
\begin{eqnarray}
&&(\Sigma_{+},\Sigma_-)=\Sigma(\cos\phi,\sin\phi), \quad
(\Sigma_{+1},\Sigma_{-1})=F\Sigma(\cos\phi,\sin\phi),  \notag \\
&& (\Sigma_{+2},\Sigma_{-2})=G\Sigma (\cos\phi,\sin\phi), \quad
N=B=\Omega_{\Lambda}=0, \quad \phi=\text{constant},
\end{eqnarray}
and $Q_2$ is given by the constraint equation.

We consider the following perturbation: 
\begin{equation*}
(Q,\Sigma ,F,G)=(-3+x,1+y,-3+z,18+w).
\end{equation*}%
Close to the Kasner circle, the centre manifold can be parametrised by the
linear combination defined by the variable 
\begin{equation*}
X=\left( \frac{14+\chi }{72}\right) x-\left( \frac{\chi -4}{6}\right)
y+\left( \frac{5\chi -2}{72}\right) z+\left( \frac{2+\chi }{72}\right) w,
\end{equation*}%
and the equations of motion on the centre manifold turn into 
\begin{equation*}
X^{\prime }=X^{2}+\mathcal{O}(X^{3}).
\end{equation*}%
This implies that close to $X=0$, $X$ is always increasing. In essence, the
Kasner circle acts as a past attractor for some orbits, while for others it
is unstable to the past.

\begin{figure}[tbp]
\includegraphics[width=8cm]{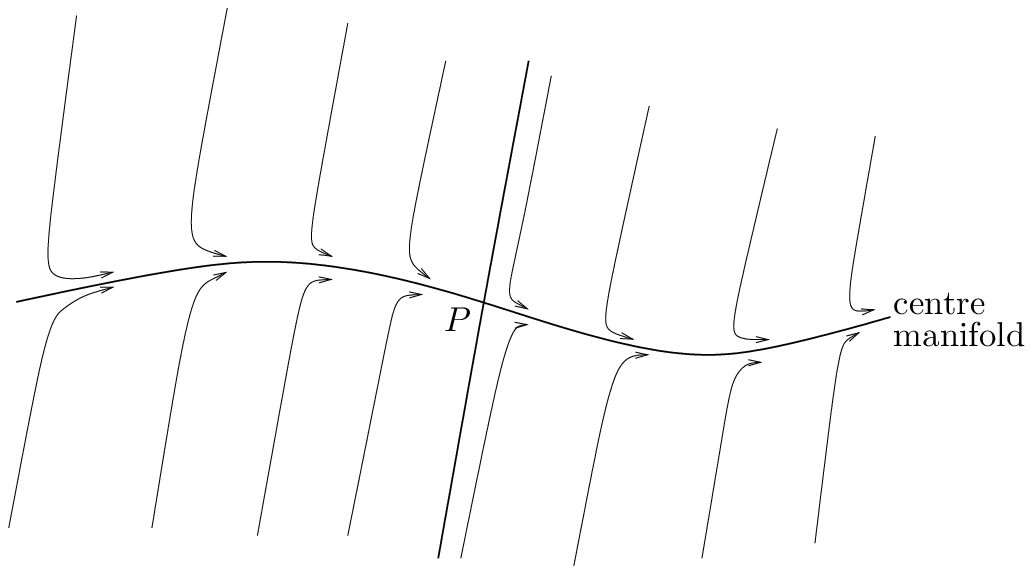}
\caption{The dynamical effect of a centre manifold: the solutions approach a
typical centre manifold and move along the centre manifold towards, or away
from, the equilibrium point $P$.}
\label{Fig:Centre}
\end{figure}

The centre manifold has a special importance for solutions asymptoting to
the equilibrium point. Due to the zero-eigenvalue the decay of the solutions
as they asymptote to the equilibrium point will be power-law (compared to
exponential) in the dynamical time, $\tau $. Hence, at sufficiently early
(or late) time, the decay rates will be dominated by the power-law
behaviour. In the phase space this can be illustrated as follows. The
solutions will approach the centre manifold exponentially rapidly, and then
move along the centre manifold towards the equilibrium point in a power-law
manner (see Fig \ref{Fig:Centre}). The centre manifold will therefore
dominate the behaviour at sufficiently early (or late) times.

A comment on the results of ref. \cite{cotsakis} is in order. They also
consider the stability of the Kasner solutions in quadratic theories of
gravity, however, they conclude that the Kasner solutions are unstable.
Technically, this is correct and they base their result on the presence of a
logarithmic term in $t$. However, as we have seen here, there exist typical
orbits having the Kasner solutions as a past attractor (in ref. \cite%
{cotsakis} this amounts to choosing a negative $k$ in their eq.(30) which
would cause the singularity to appear for $t=k>0$ and thus $\ln t$ would be
finite there). The cause of this problem is the centre manifold and it is
our opinion that the dynamical system approach unravels the true nature of
the (in)stability of the Kasner solutions more clearly. We therefore
conclude that: \emph{in the space of solutions, there exists a set of
non-zero measure which has the Kasner solutions as a past attractor}.

\subsection{Quasi-FRW solution: $\mathcal{I}$}

This solution is given by 
\begin{equation*}
\Sigma _{\pm }=\Sigma _{\pm 1}=\Sigma _{\pm 2}=N=B=\Omega _{\Lambda
}=0,\quad Q=-2,\quad Q_{2}=8.
\end{equation*}%
Its stability is determined by the eigenvalues 
\begin{equation*}
1[\times 3],\quad 2[\times 2],\quad 3[\times 3],\quad 4[\times 2].
\end{equation*}%
The presence of all positive eigenvalues (ie unstable to the future) here
implies that the solution is a past attractor.

The metric corresponding to this equilibrium point is 
\begin{equation*}
\mathrm{d}s^{2}=-\mathrm{d}t^{2}+t(\mathrm{d}x^{2}+\mathrm{d}y^{2}+\mathrm{d}%
z^{2}),
\end{equation*}%
and has $\Phi _{\mu \nu }=0,$ as can be checked explicitly -- it is the $k=0$
subcase of the radiation-like solution of eq. (\ref{quadrad}) and evolves
like a flat radiation-dominated GR universe despite being a solution of the
pure quadratic theory.

\subsection{Kasner-like solutions: $\widetilde{\mathcal{K}}(I)$}

There is also a set of Kasner-like solutions when $\chi <-1/2$: 
\begin{eqnarray}
&&Q=-1,\quad (\Sigma _{+},\Sigma _{-})=\Sigma (\cos \phi ,\sin \phi ),\quad
(\Sigma _{+1},\Sigma _{-1})=-\Sigma (\cos \phi ,\sin \phi ),  \notag \\
&&(\Sigma _{+2},\Sigma _{-2})=2\Sigma (\cos \phi ,\sin \phi ),\quad \Sigma
^{2}=\frac{1+\sqrt{-2\chi }}{-(1+2\chi )},\quad \Omega _{\Lambda }=B=N=0.
\end{eqnarray}%
These equilibrium points have both both negative and positive eigenvalues
and are therefore unstable.

The metric corresponding to this exact solution of $\Phi_{\mu\nu}=0$ is 
\begin{eqnarray}
\mathrm{d} s^2&=&-\mathrm{d} t^2+t^2\left[t^{-4\sigma_+}\mathrm{d}
x^2+t^{2(\sigma_++\sqrt 3\sigma_-)}\mathrm{d} y^2+t^{2(\sigma_+-\sqrt
3\sigma_-)}\mathrm{d} z^2\right], \\
&& (\sigma_+^2+\sigma_-^2)=\frac{1+\sqrt{-2\chi}}{-(1+2\chi)}.  \notag
\end{eqnarray}

\subsection{Super-inflating FRW universe: ${\ E}$}

This solution is given by 
\begin{equation*}
Q=Q_{2}=\Sigma _{\pm }=\Sigma _{\pm 1}=\Sigma _{\pm 2}=N=\Omega _{\Lambda
}=B=0.
\end{equation*}%
Its stability is determined by the eigenvalues: 
\begin{equation*}
0[\times 2],\quad -1,\quad -3[\times 3],\quad -\frac{3}{2}\left( 1\pm \sqrt{%
1+\frac{8(4-\chi )}{9\chi }}\right) [\times 2].
\end{equation*}%
The exact solution corresponds to an inflating Einstein metric $G_{\mu \nu
}=\lambda g_{\mu \nu }$ with $\lambda $ arbitrary; hence it is necessarily
also a solution to $\Phi _{\mu \nu }=0$.

However, interestingly, this solution has a non-trivial centre manifold
which causes a peculiar phenomenon. On the 2-dimensional centre manifold we
can use $B=X$ and $\Omega _{\Lambda }=Y$ as coordinates. To lowest order in
its neighbourhood we get 
\begin{equation*}
X^{\prime }\approx \frac{1}{6}X^{2},\quad Y^{\prime }\approx \frac{1}{6}XY.
\end{equation*}%
So if $(4-\chi )/\chi <0$, $B<0$, this solution is stable, but is otherwise
unstable.

We can show that, by approximating the solution close to the equilibrium
point, and assuming that the solution is stable, that the evolution of the
scale factor is given by $a(t)\propto e^{H_{0}t^{2}/2}$ where $t$ is the
cosmological time. The Hubble scalar diverges linearly $H=H_{0}t$ and there
are divergent curvature modes with $R\sim t^{2}$ as $t\rightarrow \infty $.
Hence, generic solutions with $B<0$ approaching ${E}$ will have deceleration
parameter $q=-[1+1/(H_{0}t^{2})]<-1$ with $q\rightarrow -1$ at late times.
These models are therefore 'super-inflationary' (or marginally
'phantom-like').

\subsection{Anisotropically-inflating type I universes: $\mathcal{A}(I)$}

Unusually, for certain values of $\chi $ and $B$, there are also exact
solutions that describe anisotropic inflationary solutions of Bianchi type
I: 
\begin{eqnarray}
&&(\Sigma _{+},\Sigma _{-})=\Sigma (\cos \phi ,\sin \phi ),\quad \Sigma
^{2}=-\frac{2(4-\chi )+B}{4(2\chi +1)}  \notag \\
&&Q=\Sigma _{\pm 1}=\Sigma _{\pm 2}=N=0.  \notag
\end{eqnarray}%
There are two classes of such solutions, depending on the values of $B$ and $%
\Omega _{\Lambda }$: 
\begin{eqnarray}
\text{(i)} &:&B=\mathrm{constant},\quad \Omega _{\Lambda }=\frac{18\chi -B}{%
8(2\chi +1)},  \notag \\
\text{(ii)} &:&B=0,\quad \Omega _{\Lambda }=\mathrm{constant}.  \notag
\end{eqnarray}

As long as $\chi $ and $B$ take values for which $\Sigma ^{2}>0,$ these
solutions exist. Their eigenvalues can be jointly expressed as: 
\begin{equation*}
0[\times 2],\quad -3[\times 3],\quad -\frac{3}{2}\left( 1\pm \sqrt{1-\frac{2B%
}{9}}\right) ,\quad -\frac{3}{2}\left( 1\pm \sqrt{1+8\Sigma ^{2}}\right)
,\quad -(1+4\Sigma _{+}).
\end{equation*}%
Two of the zero eigenvalues appear because these are 2-parameter families of
equilibrium points. However, since $\Sigma ^{2}>0$, there will always be one
unstable mode, and hence, these solutions are saddle points.

The metrics corresponding to the case where $B\neq 0$ can be written 
\begin{eqnarray}
&&\mathrm{d}s^{2}=-\mathrm{d}t^{2}+e^{2bt}\left[ e^{-4\sigma _{+}t}\mathrm{d}%
x^{2}+e^{2(\sigma _{+}+\sqrt{3}\sigma _{-})t}\mathrm{d}y^{2}+e^{2(\sigma
_{+}-\sqrt{3}\sigma _{-})t}\mathrm{d}z^{2}\right] , \\
&&b^{2}=\frac{1+8\Lambda (\alpha +\beta )}{9\beta },\quad (\sigma
_{+}^{2}+\sigma _{-}^{2})=-\frac{1+2\Lambda (4\alpha +\beta )}{18\beta }. 
\notag
\end{eqnarray}%
This set of solutions is defined so long as $b^{2}>0$ and $(\sigma
_{+}^{2}+\sigma _{-}^{2})>0$. As we can see, these solutions inflate
anisotropically. Note that they are solutions of the theory in which $%
\Lambda >0$ but are not de Sitter spacetimes. Therefore, they show
explicitly that the usual cosmic no-hair theorem of GR \cite%
{jbhair,bou,starob,jss,wald} does not hold in these theories. In effect, the
non linear ($\Phi _{\mu \nu }\neq 0$) terms contribute an effective stress
tensor to the vacuum equations which violates the strong-energy condition
needed for the cosmic no-hair theorem to hold in GR \cite{jb4}. The
essential features of this novel solution arise because of the contribution
of the quadratic Ricci terms. If we put $\alpha =0$ then there is no
essential change in the character of the solution but note that then
solution does not have a GR limit when we take $\beta \rightarrow 0$: it is
an intrinsically quadratic solution of the higher-order theory.

\subsection{Anisotropically inflating type II universe: $\mathcal{A}(II)$}

The phenomenon of anisotropic inflation is not confined to the Bianchi type
I models. An anisotropically inflating solution of Bianchi type II is
described by the critical point

\begin{eqnarray}
&&\Sigma_+=-\frac 14, \quad \Sigma_{-}=\Sigma_{\pm 1}=\Sigma_{\pm 2}=Q=Q_2=0,
\notag \\
&& \Omega_{\Lambda}=\frac{33}{32}-\frac{1}{2}N^2, \quad
B=4(1+8\chi)N^2-\frac 34(11-2\chi).
\end{eqnarray}
This is the type II solution given in \cite{BHII}.

Its stability is determined by the eigenvalues 
\begin{equation*}
0,\quad -\frac{3}{2}\left( 1\pm \sqrt{1-\frac{2B}{9}}\right) ,\quad -\frac{3%
}{2}\left( 1\pm \frac{1}{6}\sqrt{a\pm \sqrt{b}}\right) ,\quad -3,\quad -%
\frac{3}{2}\left( 1\pm \sqrt{1+16N^{2}}\right),
\end{equation*}%
where 
\begin{equation*}
a=15(3-16N^2), \quad b=9(9-6240N^2-24320N^4).
\end{equation*}
The zero eigenvalue corresponds to the fact that this solution can be
parametrised by $B$. Due to the presence of a positive eigenvalue, this
solution is unstable to the future.

There is a related set of equilibrium points for which 
\begin{eqnarray}
&&\Sigma _{+}=-\frac{1}{4},\quad \Sigma _{-}=\Sigma _{\pm 1}=\Sigma _{\pm
2}=Q=Q_{2}=B=0,  \notag \\
&&N^{2}=\frac{3}{16}\left( \frac{11-2\chi }{1+8\chi }\right) ,\quad \Omega
_{\Lambda }=\mathrm{constant}.
\end{eqnarray}%
These equilibrium points are defined for all $\chi ,$ as long as $N^{2}>0$.

\subsection{Other solutions}

We must stress that this list of equilibrium points of type II is not
exhaustive. There are further critical points of the full dynamical system,
which we do not consider here.

\section{Behaviours at infinity}

\subsection{FRW case}

Consider the behaviour at infinity for the FRW case; i.e., with $\Sigma
_{\pm }=\Sigma _{\pm 1}=\Sigma _{\pm 2}=N=0$. We note that for large values
of $Q$ there are solutions which, to lowest order, give (noting that, after
performing a translation of time, we can assume that $Q$ diverges at $\tau
=0 $ without loss of generality): 
\begin{equation}
Q=\frac{1}{2\tau },\quad Q_{2}=\frac{Q_{2,0}}{|\tau |^{\frac{3}{2}}},\quad B=%
\frac{B_{0}}{|\tau |},\quad \Omega _{\Lambda }=\frac{\Omega _{\Lambda 0}}{%
|\tau |},  \label{FRWinfinity}
\end{equation}%
where $Q_{2,0}$, $B_{0}$, $\Omega _{\Lambda ,0}$ are constants fulfilling
the condition 
\begin{equation*}
B_{0}\Omega _{\Lambda ,0}=-\frac{1}{2}.
\end{equation*}%
Hence, for these solutions at infinity to exist, the constants $B_{0}$ and $%
\Omega _{\Lambda ,0}$ need to have opposite sign; so, since $\Omega
_{\Lambda }>0$, we must have $B_{0}<0$.

Therefore, these behaviours at infinity come in two classes, according as to
whether $\tau <0$, which implies $Q<0$ and $Q$ diverges into the future, or $%
\tau >0$, which implies $Q>0$ and $Q$ diverges to the past. These two
choices for the sign of $Q$ correspond to \emph{recollapsing} and \emph{%
bouncing} cosmological solutions, respectively.

Consider the case $\tau <0$; by direct integration, we get for the Hubble
scalar and the proper time $t$: 
\begin{equation*}
H\approx H_{0}|\tau |^{\frac{1}{2}},\quad {|\tau |}^{\frac{1}{2}}\approx 
\frac{H_{0}}{2}(t_{0}-t),
\end{equation*}%
and hence, $H=(H_{0}^{2}/2)(t_{0}-t)$ and the universe reaches a maximum
size at $t_{0}$ and contracts thereafter. The divergence of the variables
for these cases is just a feature of the expansion-normalised variables.

Let us now consider the FRW case in more detail. Assuming a FRW metric, the
equations of motion can be written (without introducing expansion-normalised
variables, but assuming $(3\alpha +\beta )<0$): 
\begin{equation}
6H^{2}\dot{H}+2H\ddot{H}-\dot{H}^{2}-\eta H^{2}+\omega ^{2}=0,  \label{FRWeq}
\end{equation}%
where 
\begin{equation*}
\eta =-\frac{1}{2(3\alpha +\beta )},\quad \omega ^{2}=-\frac{\Lambda }{%
6(3\alpha +\beta )}.
\end{equation*}%
and overdots denote $d/dt$. We note that $\omega $ can have either sign and $%
\eta >0$.

We are interested in the case where there is a point where $H=0$ (i.e., the
evolution has a turning point). Let us assume that this occurs for $t=0$.
Then we can find solutions which, close to $t=0,$ can be expanded as 
\begin{equation}
H=\omega t+\frac{H_{2}}{2}t^{2}-\frac{\omega }{6}(6\omega -\eta )t^{3}-\frac{%
H_{2}}{24}(15\omega -\eta )t^{4}+\mathcal{O}(t^{5}).  \label{solH}
\end{equation}%
with $H_{2}$ constant. As eq.(\ref{FRWeq}) is ill-posed through $H=0,$ we
have assumed that $H(t)$ is analytic at $t=0$.

We note that in the case $\omega <0$, $H$ goes from being positive to
negative, hence the universe goes from an expanding to a collapsing phase.
If $\omega >0$, the universe goes from a collapsing to an expanding phase,
and so the universe experiences a bounce at $t=0$. We also note that this
turning point is not symmetric with respect to $t=0$ as long as $H_{2}\neq 0$%
. The constant $H_{2}$ is proportional to the constant $Q_{2,0}$ from above.
Moreover, by comparing the approximate solution eq.(\ref{solH}), we can
identify the behaviour at infinity, as given by eq.(\ref{FRWinfinity}), as
turning points of the expansion of the universe. Note that for the special
case $\omega =\eta =0$ the universe expands, reaches $H=0,$ and then
continues to expand.

\subsection{General case}

In the general Bianchi type I and II cases we can also find approximate
solutions at infinity, with 
\begin{eqnarray}
&&Q=\frac{1}{2\tau },\quad Q_{2}=\frac{Q_{2,0}}{|\tau |^{\frac{3}{2}}},\quad
B=\frac{B_{0}}{|\tau |},\quad \Omega _{\Lambda }=\frac{\Omega _{\Lambda 0}}{%
|\tau |},  \notag \\
&&\Sigma _{\pm }=\frac{\Sigma _{\pm ,0}}{|\tau |^{\frac{1}{2}}},\quad \Sigma
_{\pm 1}=\frac{\Sigma _{\pm 1,0}}{|\tau |},\quad \Sigma _{\pm 2}=\frac{%
\Sigma _{\pm 2,0}}{|\tau |^{\frac{3}{2}}},\quad N=\frac{N_{0}}{|\tau |^{%
\frac{1}{2}}}.  \label{eq:inf}
\end{eqnarray}%
The constants $B_{0}$, $\Omega _{\Lambda 0}$, $\Sigma _{\pm ,0}$, $\Sigma
_{\pm 1,0}$, $\Sigma _{\pm 2,0}$ and $N_{0}$ must fulfill a complicated
constraint arising from the $1/\tau ^{2}$ term of the constraint equation.

From this behaviour, we can again show that these infinities correspond to
turning points of the expansion of the universe between states of expansion
and contraction.

\subsection{Transitions at infinity}

Since the infinities described above correspond to turning points in the
evolution, we can regularly pass through these infinities by switching to,
for example, cosmological time. By choosing these infinities to occur at $%
\tau =0,$ we can pass through $\tau =0$ ($\tau <0~\mapsto \tau >0$ will
correspond to a transition from $H>0$ to $H<0$ to the future, while the
reverse will be a bounce in the past). In terms of the approximations, eq.(%
\ref{eq:inf}) we get the following transitions for the expansion-normalised
variables and the dynamical time at infinity: 
\begin{equation*}
(\tau ,Q,Q_{2},B,\Omega _{\Lambda },\Sigma _{\pm },\Sigma _{\pm 1},\Sigma
_{\pm 2},N)\mapsto (-\tau ,Q,-Q_{2},B,\Omega _{\Lambda },-\Sigma _{\pm
},\Sigma _{\pm 1},-\Sigma _{\pm 2},-N).
\end{equation*}%
These transitions can now be used to continue these solutions through the
formal infinity arising in state space. We should emphasise that for some
invariant subspaces there might be different transitions to the one given
above.

\section{Behaviour of FRW universes}

Let us consider the FRW case in further detail. It is defined by $\Sigma
_{\pm }=\Sigma _{\pm 1}=\Sigma _{\pm 2}=N=0$ and we have $\Omega _{\Lambda
}>0$. It is also useful to define the following quantity: 
\begin{equation*}
K=\frac{\Omega _{\Lambda }}{B},\quad K^{\prime }=0.
\end{equation*}%
So $K$ is a constant and we will subsequently replace all occurrences of $%
\Omega _{\Lambda }$ with $KB$. Note also that $B<0$ implies $K<0$, and $B>0$
implies $K>0$. Moreover, we will use the constraint equation to solve for $%
Q_{2}$. The equations of motion are then reduced to a two-dimensional
system: 
\begin{eqnarray}
B^{\prime } &=&-2QB, \\
Q^{\prime } &=&-\frac{3}{2}Q^{2}-3Q-\frac{1}{4}B+\frac{K}{4}B^{2},
\end{eqnarray}%
which can now be analysed by using standard techniques.

It is also useful to map the two-dimensional state space onto the compact
two-dimensional unit disk, $D^{2}$. This can be done by introducing polar
coordinates $(B,Q)=R(\cos \phi ,\sin \phi )$, followed by, for example, the
transformation: $R\mapsto R/\sqrt{1+R^{2}}$. The behaviour of the state
space at infinity is now mapped onto the unit circle $S^{1}=\partial D^{2}$.
On the unit circle there are two points of special importance, labelled by
the value of the angular variable, $\phi $: 
\begin{eqnarray}
\mathcal{B}_{\infty } &:&\quad \tan \phi =-\sqrt{\frac{|K|}{2}},\quad \phi
\in \left( \frac{\pi }{2},{\pi }\right) ,  \notag \\
\mathcal{R}_{\infty } &:&\quad \tan \phi =+\sqrt{\frac{|K|}{2}},\quad \phi
\in \left( \pi ,\frac{3\pi }{2}\right) .
\end{eqnarray}%
The points $\mathcal{B}_{\infty }$ and $\mathcal{R}_{\infty }$ correspond to
the possible turning points of the expansion (bounce and recollapse,
respectively).

There are two other points on the unit circle that correspond to $H=0$,
namely $\phi =\pi /2$ and $\phi =3\pi /2,$ which will be denoted $\mathcal{T}%
_{\infty }^{+}$ and $\mathcal{T}_{\infty }^{-}$, respectively. Both of these
points are in the invariant subspace $B=0$, but are both unstable for models
with $B\neq 0$. These two points correspond to $\omega =\eta =0$ in eq.(\ref%
{solH}).

\begin{figure}[tbp]
\includegraphics[width=8cm]{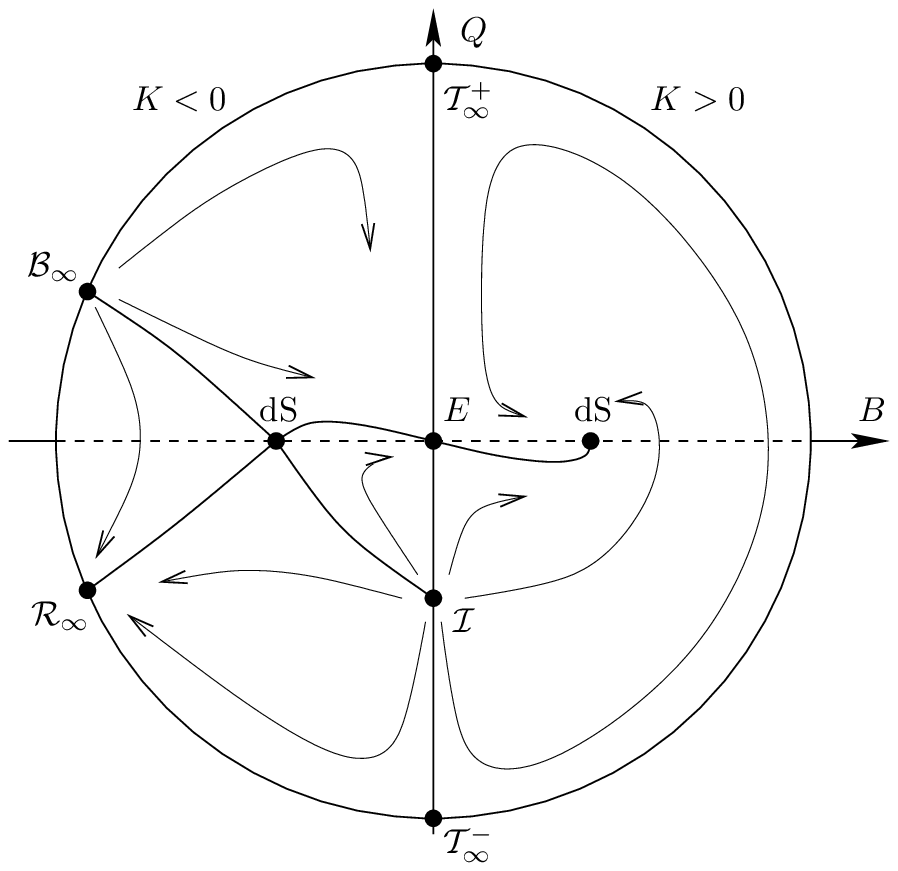}
\caption{A sketch of the dynamical behaviour of vacuum FRW universes. For
illustration, the dynamical system is mapped onto the unit disc as explained
in the text.}
\label{Fig:FRW}
\end{figure}

For $B=0,$ we can find the exact solution 
\begin{equation*}
Q(\tau )=\frac{-2}{1+Ce^{3\tau }},
\end{equation*}%
for which the metrics can be written 
\begin{equation*}
\mathrm{d}s^{2}=-\frac{k^{2}\mathrm{d}\tau ^{2}}{(1+\epsilon e^{-3\tau })^{%
\frac{4}{3}}}+e^{2\tau }\left( \mathrm{d}x^{2}+\mathrm{d}y^{2}+\mathrm{d}%
z^{2}\right) ,\quad \epsilon =-1,1.
\end{equation*}%
These metrics are solutions to $\Phi _{\mu \nu }=0$ and describe universes
evolving as follows (see Fig.\ref{Fig:FRW}):

\begin{enumerate}
\item {} $\epsilon =1$: From $\mathcal{I}$ to $E$.

\item {} $\epsilon =-1$: For $\tau <0$, from $\mathcal{I}$ to $\mathcal{T}%
_{\infty }^{-}$, while for $\tau >0$, from $\mathcal{T}_{\infty }^{+}$ to $E$%
. Note that,after requiring analyticity, we have the transition $\mathcal{T}%
_{\infty }^{-}\rightarrow \mathcal{T}_{\infty }^{+}$ at $\tau =0$.
\end{enumerate}

For $B=0$ on the approach to $E$ we have $H=H_0$ at late times unlike the
general case ($B\neq 0$) for which the evolution gives $H=H_0 t$ at late
times.

The behaviour of general FRW universes ($B\neq 0$) can be split into 5
different possibilities. The evolution can be as follows (see Fig.\ref%
{Fig:FRW}):

\begin{enumerate}
\item {} $B>0$: $\mathcal{I}~\rightarrow~ \mathrm{dS}$.

\item {} $B<0$: $\mathcal{I}~\rightarrow~ E$.

\item {} $B<0$: $\mathcal{I}~\rightarrow~ \mathcal{R}_{\infty}$.

\item {} $B<0$: $\mathcal{B}_{\infty}~\rightarrow~ E$.

\item {} $B<0$: $\mathcal{B}_{\infty}~\rightarrow ~\mathcal{R}_{\infty}$.
\end{enumerate}

The last possibility suggests that it might even be possible for a universe
to experience a sequence of expanding and collapsing phases (recall that a
collapsing phase can be described by the same equations by running the
dynamical time, $\tau $, backwards).

\section{Past behaviour of Bianchi-type universes}

In the analysis above we have seen that there are (at least) three
possibilities for past behaviour of Bianchi models. Let us discuss each of
these in turn. Again, we should emphasise that the list of equilibrium
points for the Bianchi type II case is not exhaustive, and therefore
behaviours other than these might occur in the general type II solution.

\subsection{Isotropic singularity}

We have already seen that the radiation-like FRW universe denoted $\mathcal{I%
} $, is an attractor for the models studied here. This creates the
possibility for an isotropic past singularity in these theories. Hence, the
quadratic theories of gravity studied here seem to have an isotropisation
mechanism into the past provided by the quadratic curvature contributions.
Let us give an argument that this isotropic singularity is indeed an
attractor \emph{for all Bianchi models}.

\subsubsection*{Curvature modes}

As explained earlier, the various Bianchi models can be described in terms
of the structure coefficients via the symmetric tensor $n_{ab}$ and the
vector $a_{b}$. The evolution of these curvature variables are determined by
the Jacobi identity; hence, the evolution of $n_{ab}$ and $a_{b}$ are
therefore theory-independent. Introducing expansion-normalised variables in
the standard manner, we obtain the equations of motion close to the
isotropic singularity: 
\begin{equation*}
N_{i}^{\prime }=-(Q+1)N_{i},\quad A^{\prime }=-(Q+1)A.
\end{equation*}%
Since $Q=-2$ for the attractor $\mathcal{I}$, we have 
\begin{equation*}
N_{i}\approx N_{i,0}e^{\tau },\quad A\approx A_{0}e^{\tau }
\end{equation*}%
and hence each quantity decays to the past $\tau \rightarrow -\infty $.

\subsubsection*{Shear modes}

We do expect extra shear modes to appear; however, if we introduce the
corresponding expansion-normalised variables for these modes, the evolution
equations are, to linear order, the same as for $\Sigma _{\pm }$.
Consequently, this would give eigenvalues $\{1,2,3\}$ for each of the
additional shear modes. Therefore, we expect the shear to decay as 
\begin{equation*}
\Sigma _{ij}=C_{ij}e^{\tau }+\mathcal{O}(e^{2\tau }),
\end{equation*}%
at early times \ as $\tau \rightarrow -\infty $. This is consistent with the
situation found for the Bianchi type IX model by Cotsakis et al \cite%
{cotsakis}.

\subsubsection*{Perfect-fluid matter}

With regards to the influence of matter on the evolution, let us assume the
presence of a comoving perfect fluid with a barotropic equation of state: 
\begin{equation*}
p=(\gamma -1)\rho .
\end{equation*}

Introducing an expansion-normalised energy-density, $\Omega \equiv \rho
/(3H^{2})$, the conservation of energy yields the equation of motion: 
\begin{equation*}
\Omega ^{\prime }=-[2(Q+1)+(3\gamma -2)]\Omega .
\end{equation*}%
Hence, we have for the evolution of the fluid density with volume expansion, 
$\Omega \propto e^{(4-3\gamma )\tau },$ and there is a change of behaviour
of $\Omega $ at $\gamma =4/3$. However, in the evolution equations $\Omega $
only appears as part of the product $B\Omega \propto e^{(8-3\gamma )\tau }$.
This means that for all regular matter with $\gamma \leq 2,$ the matter term
is subdominant and the isotropic singularity is stable to the past. In GR
the situation is very different, and stability is only possible when $\gamma
=2$ \cite{qui}.

These conclusions have been drawn for the case of irrotational matter only.
The situation with rotation or non-comoving 4-velocities, where $\mathbf{u}$
is not hypersurface orthogonal remains to be investigated. We know from
experience with Bianchi-type universes in general relativity that the
introduction of all the possible non-comoving velocity components can
produce a situation that is delicately sensitive to the equation of state
because of the role played by the pressure and density in the conservation
of angular momenta around the three orthogonal expansion axes, see refs \cite%
{vort,tilt}. The evolution of fluids possessing anisotropic pressures is
also interesting and is likely to play a very important role in the
late-time evolution of these universes, notably in the case where the
anisotropic fluid has vanishing trace and is accompanied by an isotropic
black-body radiation fluid. However, the way in which the quadratic Ricci
terms have been found to mimic the effect of an isotropic radiation stress
implies that the presence of a pure trace-free stress (for example a pure
magnetic field) may evolve in an influential way in these theories, just as
was found in the case of general relativity \cite{skew, mag}.

\subsection{Kasner circle}

The Kasner circle is an attractor for some orbits of type I and II and in
these models this behaviour is therefore typical. The type II modes cause
transitions between different points on the Kasner circle and since this is
an exact solution for ordinary GR we expect chaos as we go to the more
general vacuum type VIII, IX and VI$_{-1/9}^{\ast }$ models. However, even
though chaotic solutions exist, we do not know whether this chaotic set is
an attractor for typical orbits.

\subsection{Bounce}

We have seen that there are bouncing solutions for the type I and II
universes. However, even though this behaviour seems to be typical for these
models it is unclear whether they are typical for more general Bianchi
models. The bouncing solutions correspond to solutions coming from infinity
of the state space and are therefore particularly challenging to analyse.
Apart from for the FRW case, rigorous results regarding the generality of
these bouncing solutions are not known. The stability of the FRW behaviour
in the type IX case \cite{cotsakis} indicates that the closed FRW bounce
behaviour will occur there but there may be other behaviours that are stable
far from isotropy which have not yet been identified. The possibility for
bounce is closely related to the possibility of recollapsing solutions in
these theories. However, the analysis of this kind of behaviour using
expansion-normalised variables is plagued by the fact that the state
variables diverge. It does seem that recollapsing solutions are typical but
a more rigorous analysis is lacking at the present time.

\section{Future behaviour}

The future behaviour of these cosmologies shows a dependence of the
parameters of the theory. Unlike in general relativity, there does not exist
a simple no-hair theorem when $\Lambda >0$ for the vacuum case, or where
matter obeys the strong-energy condition. As discussed in our earlier paper,
ref \cite{BHII}, this situation arises because the contribution of the $\Phi
_{\mu \nu }$ stresses to the field equations (\ref{field}) can mimic the
form of a fluid with negative density and or pressures and so the
energy-conditions assumed to hold in the general-relativistic no-hair
theorems are effectively violated by the non-linear curvature terms. We
showed that there exist vacuum solutions with $\Lambda >0$ which do not
approach de Sitter at late times: they inflate anisotropically. A similar
effect can be seen in the study of Kaloper into the effects of the
Chern-Simons terms on the behaviour of a class of Bianchi type universes 
\cite{Kaloper}. However, conditions can be identified for which de Sitter is
an attractor at late times.

\subsection{De Sitter}

For de Sitter to be an attractor we need $B>0$ and $[1+2\Lambda (4\alpha
+\beta )]/\beta <0$. However, even in this case, numerical simulations
indicate that this behaviour is typical but not generic. But in the cases
where $B<0$, another late-time behaviour is possible:

\subsection{Super-inflating FRW}

For $B<0$ and $\chi <0$ or $\chi >4$ the equilibrium point $E$ is a future
attractor. Due to the presence of a non-trivial centre manifold, the
universe generically approaches this point with a power-law evolution. As
the solutions approach $E$, the universe eventually has $H(t)=H_{0}t$ and $%
q=-[1+1/(H_{0}t^{2})]$.

Lately, there has been an interest for so-called "phantom cosmologies" with $%
q<-1$ which can end in a singularity within finite time (a so-called 'big
rip'). The evolution described above has eventually $q<-1$ but approaches $%
q=-1$ sufficiently fast to avoid the 'big rip'\cite{rip}. The deceleration
parameter approaches $-1$ from below, while the Hubble scalar diverges
linearly in cosmological time. Other types of finite-time singularity are
possible in which $H$ and $\rho $ remain finite but there is an infinity of
the pressure and the acceleration \cite{sud1, sud2, sud3, sud4}. These
'sudden' singularities occur without any violation of the strong-energy
condition. It appears likely that they will occur in these quadratic
theories also. The situation in the presence of lagrangians that involved
only $R$ was discussed in ref. \cite{sud3}.

\section{Discussion}

In this paper we have provided a study of a range of anisotropic universes
in a quadratic theory of gravity involving all allowed quadratic curvature
invariants to appear in the lagrangian. We have seen that there is a
possibility for a stable (or even generic) isotropic singularity in
quadratic theories of gravity. There is a isotropic equilibrium point,
representing a FRW universe with scale factor $a(t)\propto \sqrt{t}$ that is
a solution of the \textit{vacuum} field equations of the quadratic theory,
which is stable into the past. Here, for simplicity, we have only considered
Bianchi type I and type II models, but these incorporate both shear and
3-curvature anisotropies, and an argument was given that this is also a past
attractor for more general Bianchi models. Therefore, it appears that
including quadratic terms provides us with a new mechanism for constraining
the initial singularity to be isotropic. This is reminiscent of the
situation proposed by Weyl\textbf{\ }curvature hypothesis, which envisages
some measure of the Weyl curvature playing the role of a gravitational
entropy, so it must initially be small (or zero, up to quantum corrections)
in order to provide a subsequent gravitational arrow of time \cite{penrose,
GH, BH}. However, it must not be the case that the quadratic stresses drive
the expansion towards isotropy on approach to any future singularity if the
universe is closed and recollapses.

This study raises many questions for further investigation of anisotropic
and inhomogeneous cosmologies that are more general than those considered
here, which we will report elsewhere. We have also found that the presence
of quadratic curvature terms leads to a range of new possibilities in
situations which would be expected to yield simple de Sitter inflation in
GR. The presence of a stress with $p=-\rho $ need not produce future
evolution towards the isotropic de Sitter metric locally. There is a
possibility for inflationary behaviour which is anisotropic because of the
effective stresses being contributed to the field equations by the quadratic
terms.

In ordinary GR, the general Bianchi types have been shown to possess a
chaotic behaviour as we approach the initial singularity. This chaotic
behaviour consists of a sequence of Kasner regimes connected via vacuum type
II transitions and possibly frame rotations. On a dynamical time scale, the
transitions are fairly short compared to the Kasner epochs \cite{bkl,cb,
bonn}. Therefore, since we noticed that the Kasner circle does indeed act as
a past attractor for some type I orbits, we could expect chaos to be present
for typical orbits in the more general Bianchi models. Certainly, the
chaotic vacuum type IX Mixmaster universe is a particular exact solution of
the quadratic theory and its chaotic behaviour has been discussed in the
presence of the quadratic terms \cite{cotsakis} and it remains to be seen
whether another form of chaotic evolution might arise, and be stable, far
from the isotropic closed FRW radiation-like model.

\section*{Acknowledgments}

We would like to thank Kei-ichi Maeda and Alexey Toporensky for valuable
discussions. SH was funded by an AARMS Post-doctoral Fellowship.


\begin{thebibliography}{99}
\bibitem{R2} J.D. Barrow and A.C. Ottewill, J. Phys. A 16, 2757 (1983).

\bibitem{barrcot} J.D. Barrow and S. Cotsakis, Phys. Lett. B. 214, 515
(1988).

\bibitem{barr} J.D. Barrow, Nucl. Phys. B 296, 697 (1988).

\bibitem{Inf} A.A. Starobinsky, Phys. Lett. B 91, 99 (1980); A. Guth, Phys.
Rev. D 23, 347 (1981); A.D. Linde, Phys. Lett. B129, 177 (1983); V. M\"{u}%
ller, H.-J. Schmidt and A.A. Starobinskii, Phys. Lett. B 202, 198 (1988); A.
Berkin and K.I. Maeda, Phys. Rev. D 44, 1691 (1991); S. Gottl\"{o}ber, V. M%
\"{u}ller and A.A. Starobinskii, Phys. Rev. D 43, 2510 (1991); A.A.
Starobinskii and H.-J. Schmidt, Class. Quantum Grav. 4, 695 (1987); H.-J.
Schmidt, Class. Quantum Grav. 5, 233 (1988); H.-J. Schmidt Gen. Rel. Grav.
25 87 (1993)

\bibitem{Schmidt} H.-J. Schmidt, gr-qc/0407095; V. M\"{u}ller and H. -J.
Schmidt, Gen. Rel. Grav. 17, 769 (1985).

\bibitem{BC} T. Clifton and J.D. Barrow , Class. Quant. Grav. 23, 2951(2006)

\bibitem{BC1} J.D. Barrow and T. Clifton, Class. Quant. Grav. 22, L1 (2005).

\bibitem{BC2} T. Clifton and J.D. Barrow, Phys. Rev. D 72, 103005 (2005)

\bibitem{BHII} J.D. Barrow and S. Hervik, Phys. Rev. D 73, 023007 (2006)

\bibitem{DT} S. Deser and B. Tekin, Phys. Rev. D67, 084009 (2003).

\bibitem{bach} R. Bach, Math. Zeitschrift 9, 110 (1921).

\bibitem{DS1} J. Wainwright and G.F.R. Ellis, eds., \textit{Dynamical
Systems in Cosmology}, Cambridge U.P. (1997); A.A. Coley, Dynamical Systems
and Cosmology, Kluwer, Academic Publishers (2003);

\bibitem{tilt} A.A. Coley and S. Hervik, Class. Quant. Grav. 21 (2004) 4193;
S. Hervik, R.J. van den Hoogen and A.A. Coley, Class. Quant. Grav. 22 (2005)
607; S. Hervik, R.J. van den Hoogen, W.C. Lim and A.A. Coley, Class. Quant.
Grav. 23 (2006) 845; S. Hervik and W.C. Lim, Class. Quant. Grav. 23 (2006)
3017.

\bibitem{LCD} J.A. Leach, S. Carloni and P.K.S. Dunsby, Class. Quant. Grav.
23, 4915 (2006) [arXiv: gr-qc/0603012]

\bibitem{cotsakis} S. Cotsakis, J. Demaret, Y. De Rop and L. Querella, Phys.
Rev. D 48, 4595 (1993); J. Demaret and L. Querella Class. Quantum Grav. 12,
3085 (1995)

\bibitem{topMG11} A.V. Toporensky, talk held at MG11, Berlin, Germany, 2006;
A.V. Toporensky, P.V. Tretyakov, gr-qc/0611068

\bibitem{jbhair} J.D. Barrow, Perturbations of a De Sitter Universe, In 
\textit{The Very Early Universe,} eds. G. Gibbons, S.W. Hawking and S.T.C
Siklos, Cambridge U.P., Cambridge, p.267, (1983).

\bibitem{bou} W. Boucher and G.W. Gibbons, In \textit{The Very Early Universe%
}, ed. G. Gibbons, S.W. Hawking and S.T.C Siklos, Cambridge U.P., Cambridge,
p. 273, (1983).

\bibitem{starob} A.A. Starobinskii, Sov. Phys. JETP Lett. 37, 66 (1983).

\bibitem{jss} L.G. Jensen and J. Stein-Schabes, Phys. Rev. D 35, 1146 (1987).

\bibitem{wald} R. Wald, Phys. Rev. D 28, 2118 (1983).

\bibitem{jb4} J.D. Barrow, Phys. Lett. B, 187, 12 (1987).

\bibitem{Dab} M. D\c{a}browski, astro-ph/0606574

\bibitem{qui} J.D. Barrow, Nature 272, 211 (1978)

\bibitem{vort} J.D. Barrow, Mon. Not. R. astr. Soc. 178, 625 (1977); C.B.
Collins, Comm. Math. Phys. 39, 131 (1974); G.F.R. Ellis and C.B. Collins,
Phys. Rep. 56, 65 (1979) I.S, Shikin, Sov. Phys. JETP 41, 794 (1976)

\bibitem{skew} J.D. Barrow, Phys. Rev. D. 55, 7451 (1997).

\bibitem{mag} J.D. Barrow, P. Ferreira, J. Silk, Phys. Rev. Lett. 78, 3610
(1997).

\bibitem{Kaloper} N. Kaloper, Phys. Rev. D 44, 2380 (1991).

\bibitem{rip} A.A. Starobinsky, Grav. Cosmol. 6 157 (2000); R.R. Caldwell,
Phys. Lett. B 545 23 (2002); A. Shulz and M.J. White, Phys. Rev. D 64 043514
(2001); J. Hao and X. Li, Phys. Rev. D 67 107303 (2003); G.W. Gibbons,
hep-th/0302199; S. Nojiri and S.D. Odintsov, Phys. Lett. B 562 147 (2003)
and B 571 1 (2003); P. Singh, M. Sami and N. Dadhich, Phys. Rev. D 68 023522
(2003); J. Hao and X. Li, Phys. Rev. D 68 04351 and 083514 (2003); M. D\c{a}%
browski, T. Stachowiak and M. Sydlowski, Phys. Rev. D 68 103519 (2003); P.
Elizalde and J. Quiroga, Mod Phys Lett A 19 29 (2004); P.F. Gonz\'{a}lez-D%
\'{\i}az, Phys Lett B 586 1 (2004); A. Feinstein and S. Jhingan, Mod Phys
Lett A\ 19 457 (2004); L.P. Chimento and R. Lazkoz, Phys Rev Lett 91 211301
(2003); E. Elizalde, S. Nojiri, and S.D. Odintsov, Phys.Rev. D 70 043539
(2004); L.P. Chimento and R. Lazkoz, Mod Phys Lett A 69 123512 (2004).

\bibitem{sud1} J.D. Barrow, Class. Quantum Grav. 21, L79 (2004)

\bibitem{sud2} J.D. Barrow, Class. Quantum Grav. 21, 5619 (2004)

\bibitem{sud3} J.D. Barrow and C.G. Tsagas, Class. Quantum Grav. 22, 1563
(2005),

\bibitem{sud4} S. Cotsakis and I. Klaoudatou, Preprint gr-qc/0409022

\bibitem{penrose} R. Penrose, in \textit{Theoretical Principles in
Astrophysics and Relativity}, eds. Lebovitz, N. R. Reid, W. M. \&
Vandervoort, P. O., University of Chicago Press, (1978); R. Penrose, \textit{The Road
to Reality}, Jonathan Cape, London, (2005);

\bibitem{GH} \O . Gr{\o }n and S. Hervik, Int. J. Theo. Ph. Gr. Th. Non-L.
Opt. 10, 29 (2003), \texttt{[gr-qc/0205026]}.

\bibitem{BH} J.D. Barrow and S. Hervik, Class.Quant.Grav. 19, 5173 (2002)

\bibitem{bkl} V. Belinskii, E.M. Lifshitz and I. Khalatnikov, Adv. Phys 19, 525
(1970)

\bibitem{cb} D. Chernoff and J.D. Barrow, Phys. Rev. Lett., 50, 134 (1983)

\bibitem{bonn} J.D. Barrow, Chaotic Behaviour and the Einstein Equations,
1984. In 'Classical General Relativity', eds. W. Bonnor, J. Islam and M.A.H.
MacCallum, CUP, Cambridge pp 25-41.
\end{thebibliography}
\end{document}